\documentclass[acmsmall]{acmart}
\AtBeginDocument{%
  }

\NewDocumentCommand \suite {m} {\texttt{#1}}

\NewDocumentCommand \component {m} {\texttt{#1}}
\begin{document}

\title{Applying Large-Scale Distributed Computing to Structural Bioinformatics -- Bridging Legacy HPC Clusters With Big Data Technologies Using kafka-slurm-agent}


\author{Pawel Rubach}
\email{pawel.rubach@sgh.waw.pl}
\orcid{0000-0001-5487-609X}
\affiliation{%
  \institution{Warsaw School of Economics}
  \city{Warsaw}  
  \country{Poland}
}
\affiliation{%
  \institution{University of Virginia}
  \city{Charlottesville}
  \state{Virginia}
  \country{USA}}

\renewcommand{\shortauthors}{Rubach P.}

\begin{abstract}

This paper presents the Kafka Slurm Agent (KSA), an open source (Apache 2.0 license) distributed computing and stream processing engine designed to help researchers distribute Python-based computational tasks across multiple Slurm-managed HPC clusters and workstations. Written entirely in Python, this extensible framework utilizes an Apache Kafka broker for asynchronous communication between its components. It is intended for non-expert users and does not require administrative privileges or additional libraries to run on Slurm. The framework's development was driven by the introduction of the AlphaFold protein structure prediction model, specifically, it was first created to facilitate the detection of knots in protein chains within structures predicted by AlphaFold. KSA has since been applied to several structural bioinformatics research projects, among others, leading to the discovery of new knotted proteins with previously unknown knot types. These knotted structures are now part of the AlphaKnot 2.0 web server and database, where KSA is applied to manage the knot detection process for user-uploaded structures.
\end{abstract}

\ccsdesc[500]{Computer systems organization~Grid computing}
\ccsdesc[500]{Applied computing~Bioinformatics}
\ccsdesc[300]{Information systems~Computing platforms}
\ccsdesc[100]{Software and its engineering~Software libraries and repositories}
\ccsdesc[100]{Software and its engineering~Integration frameworks}

\keywords{Distributed computing, HPC, Slurm, Kafka, Python, Bioinformatics, AlphaFold, Proteins, Protein Science, Knots, Knotted structure}

\received{8 February 2025}

\maketitle

\section{Introduction}

In the era of Big Data despite the availability of modern distributed computing platforms such as Apache Spark \cite{noauthor_spark_nodate}, Apache Flink \cite{noauthor_flink_nodate}, and even older solutions such as the SORCER platform \cite{soblewski_sorcer_2008}, the majority of computing resources, in particular, in the domain of science, are still managed by well-established centralized workload managers such as Slurm \cite{goos_slurm_2003}. While Big Data platforms are well-suited for cloud computing and can be easily set up to run on virtual machines or lightweight containers, deploying them on centralized clusters presents significant challenges. This is partly because most end users of High-Performance Computing (HPC) lack a computer science background and are accustomed to running their programs through simple bash scripts. Furthermore, many UNIX administrators prefer to manage simpler computational tasks rather than to set up complex distributed computing platforms, which often lack robust security features and can make resource management cumbersome.

A solution developed out of necessity to address these challenges is the Kafka Slurm Agent available at: \url{https://github.com/prubach/kafka-slurm-agent/} \cite{prubach_prubachkafka-slurm-agent_2025} as well as a Python package (\suite{kafka-slurm-agent}). This open source (Apache 2.0 license) distributed computing and stream processing engine, built in Python, bridges the gap between centralized clusters and independent workstations. It enables large-scale computing across various resources by adding a modern management layer with a REST API – an essential feature that legacy cluster solutions typically lack. 
This tool allows users to submit tasks and run them concurrently on multiple Slurm-managed clusters and individual workstations. The only requirement is that an Apache Kafka \cite{noauthor_kafka_nodate} broker be exposed and accessible from every cluster node or workstation. The Kafka broker acts as an intermediary between the task \component{Submitter}, the \component{ClusterAgents}, and \component{WorkerAgents} (which manage computations on clusters and workstations, respectively) and the \component{MonitorAgent}, which collects results and monitors the progress of calculations.

\section{Motivation}
Distributed computing is a diverse and evolving field, and numerous solutions have been proposed and implemented over time. The approach to scaling computing power and distributing workloads differs between industry and academia. In the commercial sector, applications are predominantly built using microservices architecture and heavily rely on cloud computing to provide the infrastructure needed to run services. This is achieved through lightweight containers or virtual machines. These containers are then managed, deployed, and scaled using open-source tools such as Kubernetes \cite{noauthor_production-grade_nodate} or Apache Mesos \cite{kakadia_apache_mesos_2015}, or through commercial solutions such as Red Hat OpenShift \cite{osborne_openshift_2018}, VMware Tanzu \cite{noauthor_vmware_tanzu}, and others.

In contrast, the scientific community relies primarily on HPC clusters managed by well-established centralized managers such as the Slurm Workload Manager \cite{slurm}. According to Morrise Jette, its lead architect, Slurm was responsible for resource management in approximately 60\% of the Top 500 HPC systems in 2023 \cite{klusacek_job_2023}. Although modern software deployment solutions, such as lightweight containers (for example, Apptainer \cite{noauthor_apptainer_nodate}), are gradually being adopted within these HPC clusters, resources are still primarily allocated and managed using resource managers like Slurm. Notably, there are ongoing efforts to bridge the gap between industry and the scientific community by integrating Slurm with Kubernetes, as seen in projects such as Slurm and/or/vs Kubernetes \cite{wickberg_slurm_2023} and Slinky \cite{wickberg_slinky_2024}.

After several years of collaborating with engineers and developing multiple distributed computing solutions (such as the SORCER platform \cite{meersman_dynamic_2009, rubach_autonomic_2009}) or using standard Java integration technologies (e.g. message queues, Java Message Service (JMS) \cite{hapner_java_1999}, etc.), I transitioned my research efforts into the field of protein science in 2018 and shifted from Java to Python. It was then that I realized how limited Python's distributed computing frameworks were in comparison to Java. However, I discovered a strong alternative to JMS in the form of the Celery Distributed Task Queue \cite{celery}.

Celery has been successfully applied to scale up the computing power necessary to process protein and RNA structures and identify non-trivial topologies. In particular, in the case of the KnotProt 2.0 database \cite{knotprot}, knots \cite{Mansfield1994}, slipknots \cite{king_identification_2007} and knotoids \cite{goundaroulis_topological_2017} in proteins are considered, and in the case of the Genus for biomolecules \cite{10.1093/nar/gkz845} the Genus measure \cite{zajac_genus_2018} in proteins and RNAs is computed to identify interesting topologies. Both databases use Celery as their backend. While Celery is a simple and reliable solution, it does not integrate well with Slurm. During the process of computing new data for these databases, frequent conflicts have been encountered with administrators of the HPC clusters utilized. The issue stemmed from the fact that Celery workers had to be kept running on several nodes for several days or even weeks since some protein structures take very long to compute and since subsequent computing tasks went through the Celery queue, from the point of view of Slurm the workers were long-running jobs that inefficiently occupied several nodes and prevented others from using the cluster.

When Jumper et al. released AlphaFold in 2021 \cite{jumper2021highly} a new challenge appeared: searching for knots in all the protein structures predicted by the AlphaFold model. To identify knots, the HOMFLY-PT and Alexander polynomials implemented in the Topoly Python library \cite{10.1093/bib/bbaa196} are computed. This process requires substantial computational resources, and the number of initially published predicted structures in the AlphaFold Database v1 (covering approximately 200,000 structures across 48 model proteomes) was roughly equivalent to the entire Protein Data Bank (PDB) \cite{bank_rcsb_nodate}. While the PDB serves as the data source for KnotProt 2.0, its structures were computed over the course of 10 years, whereas this time the goal was to compute the same data in just a few weeks. This presented a significant challenge and served as the primary motivation to develop a solution that would allow users to take advantage of multiple Slurm-managed HPC clusters, as well as individual workstations. This need ultimately led to the creation of the Kafka Slurm Agent (KSA).

\section{Architecture}

\begin{figure}[h]
  \centering
  \includegraphics[width=\linewidth]{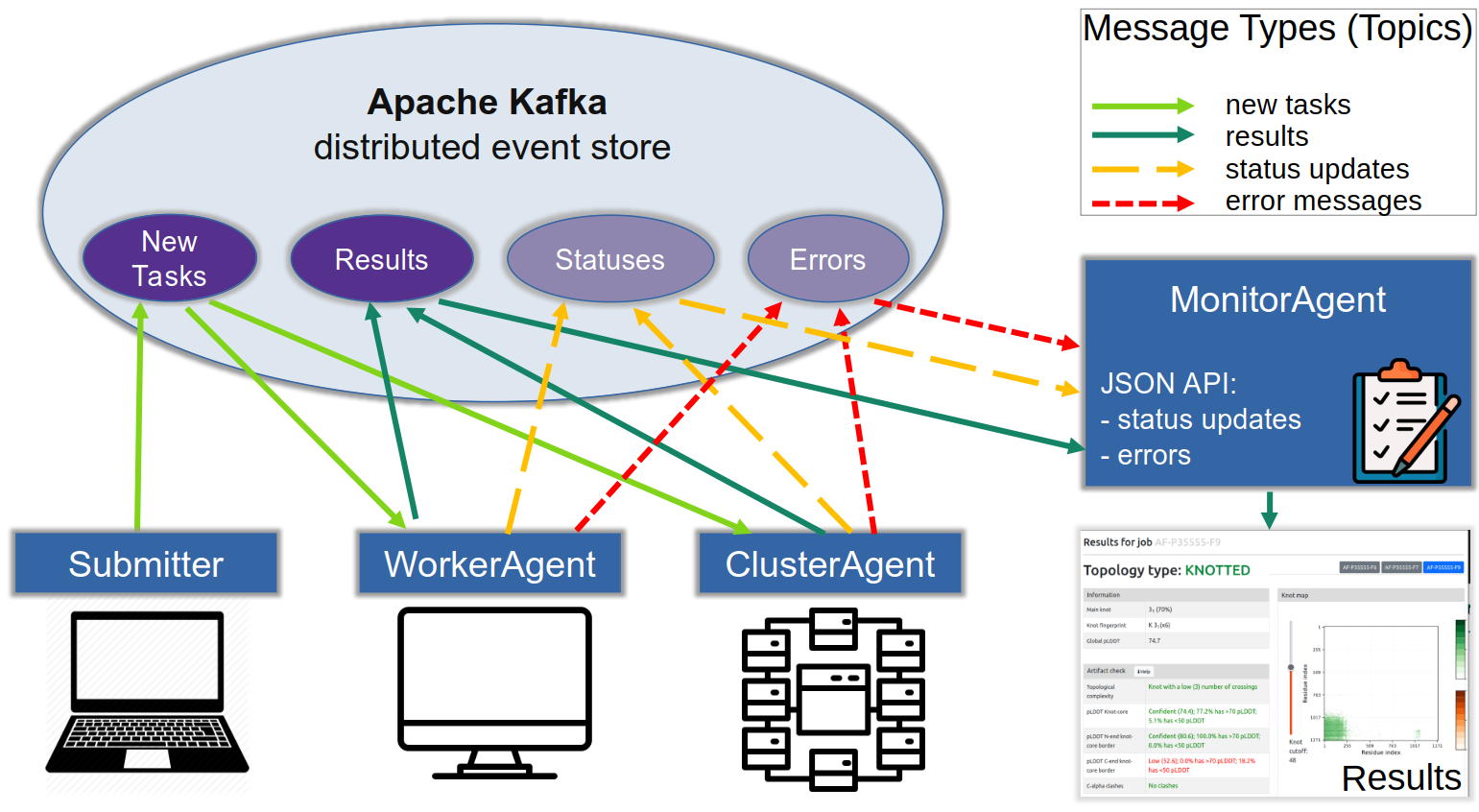}
  \caption{Kafka-slurm-agent's architecture showing components, and message flow between them.}
  \label{fig:architecture}
  \Description{Kafka-slurm-agent Architecture and components}
\end{figure}

Kafka Slurm Agent consists of four main extensible components: \component{ClusterAgent}, \component{WorkerAgent}, \component{Submitter} and \component{MonitorAgent}. The communication between all these components is asynchronous and uses the Apache Kafka broker with several topics designed to handle: new tasks, passing results, status updates, and error messages (see Fig. ~\ref{fig:architecture}).

Once the \component{Submitter} sends a new task identified by a given Task ID, this message is stored on the message broker and ready to be retrieved by the \component{ClusterAgent} or \component{WorkerAgent}. Depending on the user's available computing resources, there may be many \component{ClusterAgents}, one per each HPC cluster as well as many \component{WorkerAgents}, one per each workstation. 

The \component{ClusterAgent} is a key component of the solution. It subscribes to the new tasks Kafka topic and monitors the available computing resources on a given Slurm-managed cluster. Based on the number of CPUs or GPUs available, the \component{ClusterAgent} submits tasks as Slurm jobs to the Slurm queue and manages their execution. This includes providing updates on their current status to the status updates topic and monitoring the execution time. If a task hangs or exceeds the predefined timeout, the \component{ClusterAgent} intervenes by canceling the associated Slurm job.

The \component{WorkerAgent} operates in a similar manner to the \component{ClusterAgent}, but with a key difference: it does not rely on Slurm. Instead, it executes the retrieved tasks directly on the workstation where it is running, using separate threads for each task.

The \component{MonitorAgent} is, in fact, an optional component. Its main role is to collect the results sent by each \component{ClusterAgent} and \component{WorkerAgent} upon task completion. It also monitors the status of each submitted task, including managing error messages through a separate flow with a designated topic. To simplify user interaction, the \component{MonitorAgent} provides a web-based REST API.

One of the key reasons Apache Kafka was chosen as the communication layer is its configurability, which allows users to easily switch between exactly-once and at-least-once semantics. This feature managed through consumer group settings, supports various scenarios. For instance, one setup could involve multiple \component{MonitorAgents}, each receiving a copy of the results and status updates. Alternatively, another setup could load-balance the final processing of results between multiple \component{MonitorAgents}, ensuring that each result is retrieved and handled by only one of the active \component{MonitorAgents}.

The concept of running multiple copies of each task using several \component{ClusterAgents} or \component{WorkerAgents} should, in principle, work. However, the current implementation of the status update mechanism is not designed to handle this scenario. As a result, this could be considered a potential future extension.

Another future extension could involve creating a \component{CloudAgent} that could coordinate computing tasks on cloud-based resources. The extensibility of KSA's architecture makes it relatively easy to implement such new features.

\section{Applications in Structural Bioinformatics}

KSA has been applied in several large-scale computing projects fostering Big Data in the area of structural bioinformatics and structural biology. KSA was initially developed to identify knots in protein structures predicted by the AlphaFold model.

In 1994, Mansfield \cite{Mansfield1994} demonstrated that proteins can form knots. He discovered that the proteins he studied (carbonic anhydrase) contained "shallow knots," which unravel quickly with only a slight shortening of the protein chain. In 2000, Taylor \cite{taylor_deeply_2000} further advanced this understanding by showing that proteins can also contain "deep knots," which are much more stable. Since then, numerous studies have confirmed that knots in proteins can be stable and serve specific functions \cite{sulkowska_stabilizing_2008, sulkowska_conservation_2012}. The three-dimensional structures of proteins, determined through experimental methods such as X-Ray Crystallography,  Nuclear Magnetic Resonance (NMR) and Cryo-Electron Microscopy, are stored in the PDB.  There is, however, a huge discrepancy between the number of proteins known by sequence and gathered in the UniProt database which as of January 2025 amounts to 254 million and those for which their 3D structure has been determined via experiments. As of January 2025 the 230 thousand models contained in PDB represent only around 70 thousand proteins from UniProt. Therefore, already in 1994 the Critical assessment of methods of protein structure prediction (CASP) \cite{moult_large-scale_1995} initiative started experimenting with protein structure prediction. A breakthrough came in CASP 14 (2020) when DeepMind's AlphaFold \cite{jumper2021highly} significantly outperformed earlier models and for the first time achieved accuracy that enabled the predicted structures to be of comparable quality to the ones resolved using experimental techniques. The structures generated by this model were collected in the AlphaFold Database \cite{noauthor_alphafold_nodate}.

\begin{figure}[h]
  \centering
  \includegraphics[width=\linewidth]{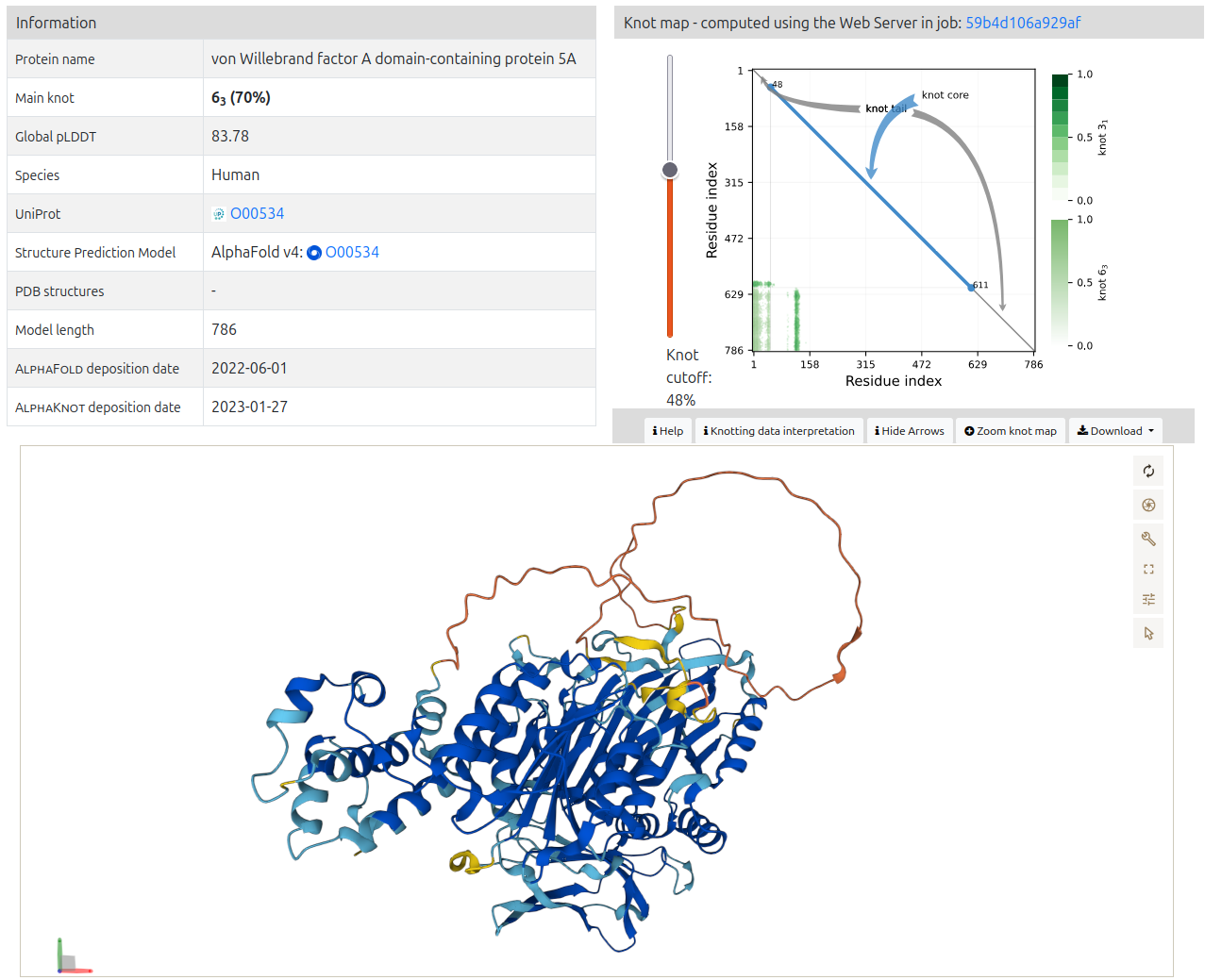}
  \caption{AlphaKnot 2.0 showing the "von Willebrand factor A domain-containing protein 5A" from the human proteome, UniProt ID: O00534, with the main knot of type $6_{3}$, \url{https://alphaknot.cent.uw.edu.pl/view/O00534/}}
  \label{fig:alphaknot}
  \Description{AlphaKnot 2.0 showing the "von Willebrand factor A domain-containing protein 5A" from the human proteome, UniProt ID: O00534, with the main knot of type $6_{3}$}
\end{figure}

This new data prompted the search for knotted proteins among the predicted structures. The first round of computations involved scanning the entire human proteome, which included 23,391 structures. This led to the discovery of the most complex knot found so far in human proteins, a six-crossing knot (type $6_{3}$) \cite{perlinska_alphafold_2023} (see Fig. \ref{fig:alphaknot}). The next stage focused on 20 model proteomes released by AlphaFold in the fall of 2021. 
The results of this search were compiled into a newly created web server and database called AlphaKnot \cite{10.1093/nar/gkac388}. For each structure, knot detection was first performed using the HOMFLY-PT polynomial on the entire chain. For structures with a probability of being knotted greater than 0.42 and a probability of being unknotted less than 0.5, a full knot map \cite{sulkowska_conservation_2012} was generated using the Alexander polynomial.
This process required significantly more computational resources, as the polynomial invariant was calculated for all subchains of the main protein chain, created by cutting subsequent numbers of amino acids off both the C- and N-ends of the protein.
The knot map enables precise identification of the position of the knot (the "knot core") within a protein chain, and thus allows for a distinction between "shallow knots" and "deep knots". For this purpose, the GPU implementation of the Alexander polynomial from the Topoly library \cite{10.1093/bib/bbaa196} was used. These calculations were run on two HPC clusters and took several weeks to complete.

Another important step was the release of the AlphaFold Database version 3 in the fall of 2022 which for the first time contained predicted structures of all proteins from the UniProt database (214 million at that time). Some of these structures (about 9 million) were later updated to version 4. This breakthrough increased the scale of analyzed data from approximately 200,000 structures, of which 5,983 were found to contain knots, to over 200 million, 681,000 of which are knotted and have been included in the new version of the AlphaKnot database known as AlphaKnot 2.0 \cite{10.1093/nar/gkae443}.

The computation of this data was a major challenge and would not be possible without Kafka Slurm Agent or a similar tool. Of the 214 million structures, around 54 million were eliminated due to low prediction quality (predicted Local Distance Difference Test (pLDDT) score for the whole chain lower than 70). For the remaining 160 million, the HOMFLY-PT polynomial was computed first with 200 random closures (the knot theory and knot polynomials are defined on closed structures, so since a protein chain is an open figure, it must be randomly closed first, which leads to the results being probabilistic: they represent the percentage of randomly closed structures that revealed a given type of knot). During the second round of calculations, the number of random closures was raised to 500 to improve the quality of knot detection. Due to the immense scale, the computation of the full knot map was not possible; instead, to determine the position of the knot, a new heuristic algorithm was developed and used. 

To complete these computations, the entire set of AlphaFold structures was divided into batches of 4,000, with each batch submitted as a single task to the \component{ClusterAgents} and \component{WorkerAgents} via KSA. The first round of computations alone took approximately three weeks to complete, utilizing three Linux HPC clusters with a total of 70 nodes and around 2,000 cores.

AlphaKnot 2.0 also includes structures generated using the Evolutionary Scale Modeling (ESM) model developed by Facebook Research and first published in November 2022 \cite{lin_evolutionary-scale_2023} which, unlike AlphaFold, is a Large Language Model.   However, since ESM did not publish structures from the UniProt database, the structures shown in AlphaKnot 2.0 were generated using the ESMfold model version 1 for proteins that were first determined to have knots using the AlphaFold model. The limitations resulting from available GPU resources, in particular memory, enabled the generation of only protein structures no longer than 400 amino acids.
The analysis of the ESM-generated protein structures collected in AlphaKnot 2.0 shows that the first version of ESM did not reach the structure prediction performance of AlphaFold v2, however, the recently published v3 \cite{hayes_simulating_2025} promises a much higher prediction accuracy.

The data computed with the help of KSA and stored in the AlphaKnot 2.0 database has contributed to further research, including a deeper understanding of knotted proteins, their occurrence across various organisms, and the roles they play \cite{perlinska_everything_2024}. It has also been used as training data for the development of machine learning models capable of identifying knotted proteins \cite{sikora_knot_2024}.

In addition to preparing the data for AlphaKnot, KSA is integrated with the application's built-in web service, which allows users to process their own structures. It manages all user requests and performs the necessary computations behind the scenes.

\section{Implementation}

KSA is designed to be as easy as possible for an unprivileged user of an HPC cluster or a workstation, and thus, these factors influence the choice of underlying technologies and libraries. It is implemented entirely in Python (currently supports versions 3.6-3.13) and relies on the \suite{kafka-python-ng} \cite{barnhart_wbarnhakafka-python-ng_2025} library for communication with the Kafka broker and \suite{faust-streaming} \cite{noauthor_faust-streamingfaust_2025} to synchronize task queues, status updates, and result data between its components. The communication with Slurm occurs without the need for any additional Slurm components such as the Slurm REST API \cite{noauthor_slurm_rest_nodate}, Slurm Job Completion Kafka plugin \cite{noauthor_slurm_jobcomp_kafka_nodate} or the Slurm C library. 

KSA is available as the \suite{kafka-slurm-agent} package in the standard PyPi repository \cite{noauthor_pypi_nodate} and can be installed using the \suite{pip} installer. After installation, the built-in \suite{kafka-slurm} command can be used to generate a runnable project template. A ready to run demonstration project "KSA Demo" is also made available on Github at: \url{https://github.com/ilbsm/ksa_demo} \cite{noauthor_ilbsmksa_demo_2025}.

\begin{figure}[h]
  \centering  
  \includegraphics[width=0.6\linewidth]{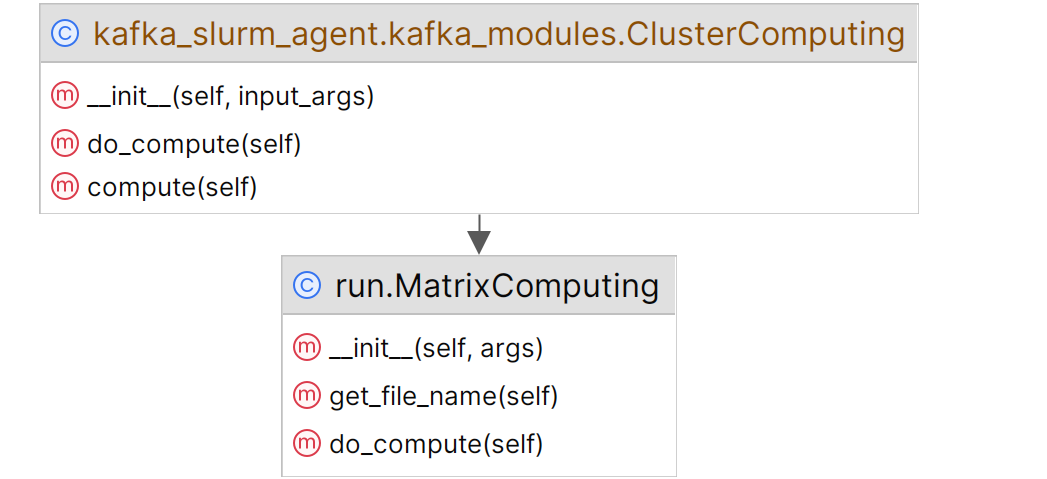}
  \caption{UML class diagram showing the basic ClusterComputing class and its methods along with an example of a user-defined MatrixComputing class that implements the script that will be run by KSA.}
  \label{fig:cluster_comp}
  \Description{}
\end{figure}

KSA allows one to run any kind of computation from a simple Python script. The script has to contain a class that extends the built-in \suite{ClusterComputing} class (Fig. \ref{fig:cluster_comp}). During the submission of tasks, one can pass several parameters. The main one is the name of the script to run. Another one is the task ID – the identifier of a particular task. Other parameters control the resources necessary for the computation: GPU, memory, and number of CPUs. Apart from all those, it is possible to pass any number and type of other parameters –- these will be serialized in the Kafka message and then passed to the computing script in the form of a JSON file and then read and made available as configuration parameters of the task.

The submission of any task involves setting the necessary parameters and then using the built-in \component{Submitter} class to send the appropriate messages, which are added to the new tasks topic on the Kafka broker. Once the tasks are submitted, they are ready to be picked up by the cluster or worker agents. These agents can then be started on Slurm-managed clusters or individual workstations, respectively.

There are four different Kafka topics used in the system. While the names of these topics can be configured in the configuration file, they are, by default, given a common prefix with suffixes that indicate their specific role in the task management process:
\begin{itemize}
\item {\texttt{PREFIX-new}}: this is the topic to which messages containing descriptions of tasks to be computed are sent and then read by cluster and worker agents.
\item {\texttt{PREFIX-jobs}}: this topic contains the statuses of particular tasks – typically those may have the following values: SUBMITTED, WAITING, RUNNING, DONE. Other values can be added and computing scripts can also send status updates at any moment of the computing process.
\item {\texttt{PREFIX-done}}: the results of finished tasks are sent to this topic.
\item {\texttt{PREFIX-error}}: in case an error occurs during the computation of a given script a message is sent to this topic.
\end{itemize}

The PREFIX and all other parameters can be set in the \suite{kafkaslurm\_cfg.py} config file.
A template project generated using the \suite{kafka-slurm} command will contain a template of the config file, \suite{Bash} scripts to start all necessary components of the solution (\component{ClusterAgent}, \component{WorkerAgent} and \component{MonitorAgent}) as well as templates of a script to be run using KSA and a script that implements the submission of tasks.

KSA should run on any operating system supporting Python 3.6 or later, however, currently, the template provides startup scripts only in \suite{Bash}, therefore if the agents were to run on Windows, an installation of \suite{Bash} or Windows Subsystem for Linux (WSL) would be necessary, or the user could easily create \suite{Batch} scripts based on the included ones in \suite{Bash}.

\section{Performance}

The performance of KSA has not been specifically measured, as it is highly case-specific, making it difficult to provide objective results. It depends on several factors:
\begin{itemize}
\item The performance of the Kafka broker and the connection between it and the KSA's Agents.
\item The number of Topic partitions configured in the Kafka broker, in particular, for the new tasks topic.
\item Cluster and Worker Agents rely on a polling mechanism to check for new tasks to start, and the polling interval is configurable. Therefore, the delay between task submission and its execution by the \component{WorkerAgent} or submission to the Slurm queue by the \component{ClusterAgent} depends largely on this parameter. The \component{ClusterAgent}'s standard strategy is to always submit more tasks to Slurm than can be immediately started, based on available resources. This approach ensures that the Slurm queue always has tasks waiting, allowing Slurm to start subsequent tasks as soon as resources become available.
\item For projects with tens of thousands of tasks or more, synchronizing task status updates may slow down the startup of KSA agents. The Apache Kafka broker handles the parameters that determine the message retention policy and cleaning within each topic.
\end{itemize}

The overhead associated with task management, status updates, and other processes makes KSA unsuitable for scenarios where each task executes within milliseconds. However, the same limitation applies to Slurm. On the other hand, for longer-running tasks, the advantages of using KSA — such as simplified task submission and monitoring -- are likely to outweigh the performance loss caused by overhead.

\section{Discussion and other solutions}

There are many approaches to distributing and parallelizing research computing. 
The Beowulf cluster \cite{noauthor_beowulf_2001, fischer_roots_2014} created in 1994 marks the beginning of modern HPC architecture -- clusters composed of multiple nodes -- inexpensive computers, each running a separate instance of UNIX-like operating systems (Linux, usually). This architectural shift necessitated the development of workload managers capable of allocating resources and scheduling. The most widely used ones today include Slurm \cite{goos_slurm_2003}, Sun Grid Engine (later renamed Oracle Grid Engine) \cite{gentzsch_sun_2001}, and Portable Batch System (PBS) Professional \cite{jones_pbs_2001}. A natural progression from this was the creation of higher-level solutions enabling the sharing of computing resources across different locations or organizations, leading to the rise of Grid Computing. Notable projects that emerged from this development include UNICORE \cite{erwin_unicore_2001} and the now-retired Globus Toolkit \cite{foster_globus_1999}. However, Grid Computing’s centralized architecture lacked high availability and reliability, prompting the development of decentralized solutions. These included peer-to-peer architectures (e.g., Sun Microsystems' Jini \cite{kumaran_jini_2001}), space-based computing (e.g., Javaspaces \cite{freeman_javaspaces_1999}), and further advancements of message queues, a concept that dates back to the 1980s. Building on decentralized solutions, higher-level programming (metaprogramming) environments were developed (e.g., SORCER \cite{soblewski_sorcer_2008}) which implemented the Service-Object Oriented Architecture \cite{sobolewski_service-oriented_2014}. Meanwhile, message queues were adopted by Sun Microsystems and turned into the Java Message Service (JMS) standard \cite{hapner_java_1999} and ultimately formed the conceptual basis for the technology known today as Apache Kafka \cite{noauthor_kafka_nodate}, created first in 2010 as an internal project at LinkedIn.

Two other key concepts that shaped modern distributed computing emerged in the 2000s. First, the ideas behind MapReduce \cite{dean_mapreduce_2008} and Google File System \cite{ghemawat_google_2003} led to the creation of Hadoop in 2008. Second, in 2006 Amazon launched Amazon Web Services \cite{noauthor_cloud_2025}, effectively marking the beginning of cloud computing. Hadoop's successor Apache Spark \cite{zaharia_spark_2016} introduced the Resilient Distributed Dataset (RDD) \cite{zaharia_resilient_2012} and shifted much of the processing to memory. Meanwhile, the enterprise software world advanced cloud computing by introducing lightweight, extensible containers (e.g., Docker \cite{noauthor_docker_2022}, Apptainer \cite{noauthor_apptainer_nodate}) and orchestration systems such as Kubernetes \cite{noauthor_production-grade_nodate}, moving from the previously dominant Service-Oriented Architecture (SOA) towards microservices.

This brief history of distributed computing is important to highlight the wide range of solutions available today and to show how KSA fits into this broader landscape.

The key difference between KSA and the solutions mentioned above is that KSA was designed with the needs of a typical unprivileged user in mind, someone with no computer science background. As a result, it is easy to deploy on both workstations and HPC resources without requiring intervention from an administrator. Another key design feature is its compatibility with the still dominant model of resource allocation through workload managers like Slurm, which is commonly used by the scientific community.

Many of the solutions mentioned earlier do not meet these two requirements. For example, some require complex deployment with administrator assistance (e.g., UNICORE, Globus Toolkit, Hadoop, HTCondor \cite{fajardo_how_2015}), while others do not align with the resource allocation model used by the scientific community and if used on HPC clusters would have to rely on long-running workers (e.g., Apache Spark, message queues, Celery \cite{celery}).

There are also scientific workflow management systems such as Snakemake \cite{koster_snakemakescalable_2012}, Nextflow \cite{di_tommaso_nextflow_2017}, and Pegasus \cite{deelman_pegasus_2015} which can also be used for similar purposes, however, they usually require administrative assistance and are more difficult to deploy.

On the other hand, there are frameworks somewhat similar to KSA, i.a. Parsl -- the Parallel Scripting Library \cite{parsl}. Parsl project focuses on parallelizing Python functions and supports several backends including Slurm. It is particularly well-suited for cases where a single computation needs to be decomposed into smaller tasks to improve overall performance. In contrast, KSA is designed to handle thousands of simple, identical, and independent tasks, which typically differ only in their input data.

Another similar solution is SmartSim \cite{partee_using_2021} which allows Python code to be executed using HPC schedulers (Slurm among them). Unlike KSA, this library is primarily aimed at supporting machine learning libraries such as PyTorch and TensorFlow, enabling model training on HPC.

\section{Conclusions}

This paper introduced the Kafka Slurm Agent, a distributed computing and stream processing engine designed to allow researchers to distribute their Python-based computing tasks across multiple Slurm-managed HPC clusters and workstations.

KSA is an extensible framework written entirely in Python. It is intended for users with minimal or no computer science background, and it does not require administrative privileges or the installation of additional libraries to integrate with and run on Slurm.

The creation of KSA was driven by recent advancements in structural bioinformatics, particularly the introduction of the AlphaFold protein structure prediction model, which effectively increased the number of available 3D structures of proteins by three orders of magnitude. KSA was tested in several real-world scenarios, involving deployment across up to three HPC clusters simultaneously, as well as several workstations. The detection of knots in protein structures predicted by AlphaFold took several weeks to complete, ultimately leading to the creation of the AlphaKnot 2.0 web server and database.

There are many ways to leverage distributed computing and HPC resources to scale up computations. While KSA is not a universal solution for every challenge faced by the research computing community, as demonstrated by the real-world applications described, it offers a useful and relatively easy-to-use tool that facilitates the transition into the era of Big Data.

\begin{acks}
The author wishes to thank Joanna Sulkowska for the motivation to join the bioinformatics community and invaluable help in discovering the world of protein topology, Wladek Minor for revealing the details of structural biology and introduction to crystallography, Tomasz Osiński for drawing attention to the PEARC community and the encouragement to write this paper, as well as Michael W Sobolewski for valuable suggestions and comments.

ChatGPT and Grammarly were used to rephrase the text and improve its styling.
\end{acks}
\bibliographystyle{ACM-Reference-Format}
\bibliography{articles,distrib_computing}

\end{document}